\documentclass{article}
\usepackage{ijcai25}

\usepackage{times}
\usepackage{soul}
\usepackage{url}
\usepackage[hidelinks]{hyperref}
\usepackage[utf8]{inputenc}
\usepackage[small]{caption}
\usepackage{graphicx}
\usepackage{amsmath}
\usepackage{amsthm}
\usepackage{booktabs}
\usepackage{algorithm}
\usepackage{algorithmic}
\usepackage[switch]{lineno}
\usepackage{amsfonts}
\usepackage{float}
\usepackage{subfig}
\usepackage{pifont}
\usepackage{stfloats}
\usepackage{multirow}

\title{Double Distillation Network for Multi-Agent Reinforcement Learning}

\author{
Yang Zhou$^1$
\and
Siying Wang$^{2*}$\and
Wenyu Chen$^1$\and
Ruoning Zhang$^1$\and\\
Zhitong Zhao$^1$\And
Zixuan Zhang$^1$\\
\affiliations
$^1$School of Computer Science and Engineering, \\University of Electronic Science and Technology of China\\
$^2$School of Automation Engineering, University of Electronic Science and Technology of China\\
\emails
zhouy@std.uestc.edu.cn,
\{siyingwang, cwy\}@uestc.edu.cn,
\{zhangruoning, zhaozhitong, 202322080910\}@std.uestc.edu.cn
}
\begin{document}

\maketitle

\begin{abstract}
  Multi-agent reinforcement learning typically employs a centralized training-decentralized execution (CTDE) framework to alleviate the non-stationarity in environment. However, the partial observability during execution may lead to cumulative gap errors gathered by agents, impairing the training of effective collaborative policies. To overcome this challenge, we introduce the Double Distillation Network (DDN), which incorporates two distillation modules aimed at enhancing robust coordination and facilitating the collaboration process under constrained information. The external distillation module uses a global guiding network and a local policy network, employing distillation to reconcile the gap between global training and local execution. In addition, the internal distillation module introduces intrinsic rewards, drawn from state information, to enhance the exploration capabilities of agents. Extensive experiments demonstrate that DDN significantly improves performance across multiple scenarios.
\end{abstract}

\section{Introduction}
Over the past decades, collaborative Multi-agent Reinforcement Learning (MARL) has shown significant potential in various fields, including unmanned aerial vehicles~\cite{yue2022unmanned}, robotic coordination~\cite{zhou2024multirobot}, and autonomous vehicles~\cite{yadav2023comprehensive}. However, cooperative agents often have limited observation ranges and can only access partial local information, which poses significant challenges to the collaboration process, especially in environments with observational instability and decision uncertainty. These challenges significantly impede agents' training on collaborative policies.

To address these non-stationary challenges, a commonly used approach is to implement the Centralized Training with Decentralized Execution (CTDE)~\cite{kraemer2016multi} paradigm, which helps to stabilize policy changes during training. The main concept of this framework is to allow agents to access global state information for policy training during the training phase, while still using only partial observations for execution to simulate the decentralized decision-making process. This framework enables agents to effectively integrate global information and significantly alleviating the non-stationarity caused by the environment. Unfortunately, current CTDE methods primarily overemphasize the factorization process of the centralized value function $Q_{tot}$ while ensuring it adheres to the Individual Global Max (IGM) criterion. They may {\it overlook} the role of the global state information, such as VDN~\cite{10.5555/3237383.3238080}, or embed these state features {\it indirectly} as in QMIX~\cite{rashid2020monotonic} and Qatten~\cite{yang2020qatten}. In other words, the features of global state information, as well as the discrepancies between global and local observations, are not effectively utilized by the policy network and decision-making process, which may hinder the training efficiency of the cooperative policies.

Moreover, there still exists an inherent error between the centralized value function which integrates global state information, and the local utility functions used for agent decision-making, even if the centralized value function is correctly factorized. This error would be accumulated gradually as the model interacts with the environment, inevitably impacting the training of cooperative policies. Therefore, the current problem becomes: {\it how to better utilize global state information while ensuring that agents make decisions based on local observations?} Recent approaches attempt to fully separate the centralized training and distributed execution stages, where agents in the execution phase are simply tasked with mimicking or replicating the utility function from training. These approaches either augment the partially observable local features or excessively substitute them with the global state~\cite{ijcai2024p0004,zhao2022ctds,hong2022rethinking}. While theoretically eliminating the inherent error aforementioned, they fail to fully leverage the state information, still hindering the enhancement of cooperative strategy training efficiency.

In this paper, we propose a mixed-distillation framework called Double Distillation Network (DDN), to eliminate the inherent error of the global state information integration. Orthogonal to the existing methods, DDN incorporates two distinct distillation modules to separately eliminate these errors and efficiently integrate the state information. Specifically, the external distillation module adopts a leader-follower architecture, where the leader consists of a Global Guiding Network(GGN), which integrates state information into each agent's observation features through personalized fusion blocks, tailored to each individual observation. The follower represents the Local Policy Network(LPN), where each agent only includes its own local observations. DDN employs a multi-layer feature alignment knowledge distillation mechanism to reduce the gap between the leader and the follower, with the inherent error being eliminated through this isolation mechanism, allowing the follower's LPN to indirectly learn from environmental feedback. Additionally, the Internal Distillation Module incorporates an intrinsic reward that is related to global state features. The intrinsic reward is derived from the difference between a randomly initialized target network and the prediction network encoding global state information, implying that the closer the two networks are, the higher the intensity of exploration. The special design of the Internal Distillation Module directly integrates global state information into the reward signal, which effectively enhances the exploration capability of the agents, significantly improving both the efficiency and quality of cooperative policy learning. The contributions of DDN are summarized as:

\begin{itemize}
    \item We propose a novel distillation network based on a leader-follower architecture, where the GGN incorporates personalized state information, while the LPN relies only on local information to perform multi-level distillation. The proposed method can effectively eliminates the inherent error in the state fusion process within the CTDE framework.
    \item We propose an Internal Distillation Module based on the global state, inspired by curiosity-driven mechanisms. This distillation procedure evaluates the access intensity of global state features and generates corresponding intrinsic rewards to promote the exploration of the collaborative strategy.
    \item A large number of experiments on SMAC prove the superiority of DDN in both leader and follower, and also verify its advantages in multi-agent cooperation tasks.
\end{itemize}

\section{Related Works}
\subsection{Existing Approaches in MARL}
Multi-agent systems are categorized as cooperative, competitive, or mixed based on agent interactions~\cite{wong2023deep}. Cooperative MARL, with its potential in applications like path planning and autonomous driving, has garnered significant research interest in recent years~\cite{oroojlooy2023review}.

Decentralized execution is essential for fully cooperative multi-agent systems in real-world scenarios, making the CTDE paradigm crucial. Some of the typical methods are as follows. VDN~\cite{10.5555/3237383.3238080} represents a foundational value decomposition, addressing the ``lazy agent" problem by decomposing the joint action-value function $Q_{tot}$ into a sum of individual $Q$-values $Q_i$, conditioned on local observations. QMIX~\cite{rashid2020monotonic} builds on VDN by introducing a hyper-network to enforce monotonicity between $Q_{tot}$ and local value functions, incorporating global state information. QTRAN~\cite{son2019qtran} relaxes the additivity and monotonicity constraints, deriving the IGM principle. Qatten~\cite{yang2020qatten} employs a multi-head attention mechanism for value decomposition. Weighted QMIX~\cite{rashid2020weighted} reduces suboptimal joint action weights to avoid local optima. QPLEX~\cite{wang2020qplex} extends IGM by using advantage-function consistency constraints and an attention mechanism. RA3~\cite{wang2023regularization} accelerates training by framing $Q_{tot}$ updates as a fixed-point iterative task. QDAP~\cite{zhao2024qdap} improves cooperation by weighting historical trajectories of agents, addressing the impact of dead agents. QEN~\cite{wang2024enhancing} enhances collaboration through a graph neural network based on Pearson correlation coefficients of agents’ trajectory similarities.

CTDE-based methods often face challenges from cumulative inherent error, limiting agents' ability to achieve optimal policies. Additionally, inadequate exploration strategies during training can hinder learning, compromising policy robustness and adaptability. For example, Nguyen~\cite{huang2024optimistic} addresses the undervaluation of individual policies and communication complexity by employing a greed-driven approach and an incentive-based communication module to foster cooperation. GDIR~\cite{tan2023goexplore} incorporates intrinsic rewards and organizes learning into ``Go" and ``Explore" phases for continuous learning and faster adaptation. Despite these advances, achieving low-cost, reliable communication remains a significant challenge in MARL, and poorly designed rewards can lead to interference in agent learning.

To address these challenges, we propose a double distillation learning method that not only mitigates inherent error through distillation models but also leverages global state information to generate state-dependent intrinsic rewards. This approach optimizes the decision-making process of agents and enhances the overall performance of multi-agent systems.

\subsection{Knowledge Distillation in MARL}
The key idea in Knowledge distillation~\cite{hinton2015distillingknowledgeneuralnetwork} is to transfer knowledge from a large teacher model to a lightweight student model, facilitating deployment across various tasks~\cite{gou2021knowledge,wang2021knowledge,xu2024survey}, which meets the need to eliminate inherent error in MARL training. Policy distillation~\cite{rusu2015policy} first applied this concept to train smaller, more efficient networks for agent policies. While CTDS~\cite{zhao2022ctds} distills policies for decentralized execution by approximating teacher estimates (based on global observations) with partial observations. PTDE~\cite{ijcai2024p0004} uses a two-phase approach to distill personalized global information into local agent policies, enabling decentralized execution. IGM-DA~\cite{hong2022rethinking} trains a global expert and decomposes its policies into local observation-based ones through imitation learning. However, these methods are aimed at indirectly incorporating state information, and the training efficiency methods still need to be improved.

Our proposed DDN incorporates two modules: an Internal Distillation Module balancing exploration and exploitation, and an external module focusing on knowledge transfer between decisions and intermediate feature learning, which can use global state information directly and efficiently to improve training efficiency.

\section{Preliminaries}
\subsection{Dec-POMDPs}
The fully cooperative multi-agent task is modeled as a decentralized partially observable Markov decision process (Dec-POMDP) ~\cite{oliehoek2016concise}, represented as a tuple $ G = \left\langle \boldsymbol{S}, \boldsymbol{U}, P, r, Z, O, \boldsymbol{N}, \gamma \right\rangle $, where $s \in \boldsymbol{S}$ denotes the global state in the environment. At each time step $t$, every agent $i \in \boldsymbol{N} \equiv { 1, \dots, n }$ picks an action $u_i \in \boldsymbol{U}$, with $\boldsymbol{u}^t = { u_1^t, u_2^t, \dots, u_n^t }$ denoting the collective actions of all agents at time $t$. The environment’s state transition function $P(s^{t+1}|s,\boldsymbol{u}) : \boldsymbol{S} \times \boldsymbol{U} \times \boldsymbol{S} \mapsto [0, 1]$ governs state transition dynamics. Once the agents execute the joint action $\boldsymbol{u}^t$, a shared reward function $r(s^t, \boldsymbol{u}^t)$ is feedback to them, where $\gamma \in [0, 1]$ is the discount factor. The goal of this cooperative team is to derive a joint policy $\pi$, which is based on the action-value function: $Q^\pi(s^t, \boldsymbol{u}^t) = \mathbb{E}_{s_{t+1:\infty},\boldsymbol{u}_{t+1:\infty}}\Big[\sum_{k=0}^\infty \gamma^k r^{t+k} \mid s^t, \boldsymbol{u}^t\Big]$.

\subsection{Value decomposition and IGM condition}
The value decomposition under the CTDE framework primarily involves decomposing a task originally completed collaboratively by multiple agents into individual tasks for each agent. The core idea is to allow each agent to focus on its own responsibilities without considering the impact of other agents. At this point, the value decomposition process involves decomposing the centralized value function, updated by the team reward from the environment, into individual agents. This process must satisfy the following equation:

\begin{equation}\label{qtot}\small
    \begin{aligned}
        Q_{tot}(s^{t}, \mathbf{u}^{t}) = & r + \gamma \max_{\mathbf{u}^{t+1}} Q_{tot}(s^{t+1}, \mathbf{u}^{t+1}) \\
        = & r + \gamma \max_{\mathbf{u}^{t+1}} \mathcal{F}( Q_{1} \left(\tau_{1}^{t}, u_{1}^{t}\right), \dots, Q_{n} \left(\tau_{n}^{t}, u_{n}^{t}\right);s^{t+1})
    \end{aligned}
\end{equation}

\noindent here $\mathcal{F}$ denotes the mixing network, responsible for decomposing the $Q_{tot}$ into individual utility functions for each agent. The agent-specific utilities $Q_i$ are updated by leveraging the trained joint value function $Q_{tot}$, with the mixing network facilitating this process. For this process to proceed, it is necessary to ensure that the joint optimal action of $Q_{tot}$ and the local optimal action of $Q_i$ align, implying that the {\bf IGM} (Individual-Global-Max) condition must hold:

\begin{equation}
    \arg\max_{u}Q_{tot}(\tau,u)=
    \begin{pmatrix}
    \arg\max_{u_1}Q_1(\tau_1,u_1) \\
    \vdots \\
    \arg\max_{u_n}Q_n(\tau_n,u_n)
    \end{pmatrix}
\end{equation}

VDN~\cite{10.5555/3237383.3238080} and QMIX~\cite{rashid2020monotonic} are classic examples of methods that satisfy the IGM condition. VDN satisfies the IGM condition by additivity $Q_{tot}(\tau,u)=\sum_{i=1}^{n}Q_{i}(\tau_{i},u_{i})$. While QMIX~\cite{rashid2020monotonic} satisfies the IGM condition via monotonicity $Q_{tot}(s,u)=f_{s}(Q_{1}(\tau_{1},u_{1}),\ldots,Q_{n}(\tau_{n},u_{n}),s),\frac{\partial f_{i}}{\partial Q_{i}}\geq0,i\in[1,n]$. These methods attempt to factorize $Q_{tot}$ assuming additivity and monotonicity aforementioned. 

\subsection{Knowledge Distillation}
KD is a popular model compression technique where a lightweight student model is trained using supervision from a powerful teacher model. The teacher’s output serves as ``knowledge", and the student learns to transfer this knowledge through ``distillation". In Computer Vision and Natural Language Processing, a high-performance teacher model with strong generalization and feature processing abilities is chosen, and the student model learns by minimizing the output difference. In the MARL community, KD can transfer agents' behaviors, policies, or Q-value functions through centralized training. The teacher agent, trained with global state information, guides the decentralized student agents, which use only local observations. This approach aims to reduce errors in both the centralized value function and local utility function in the proposed DDN.

\begin{figure*}[t]
    \centering
    \includegraphics[width=0.83\textwidth]{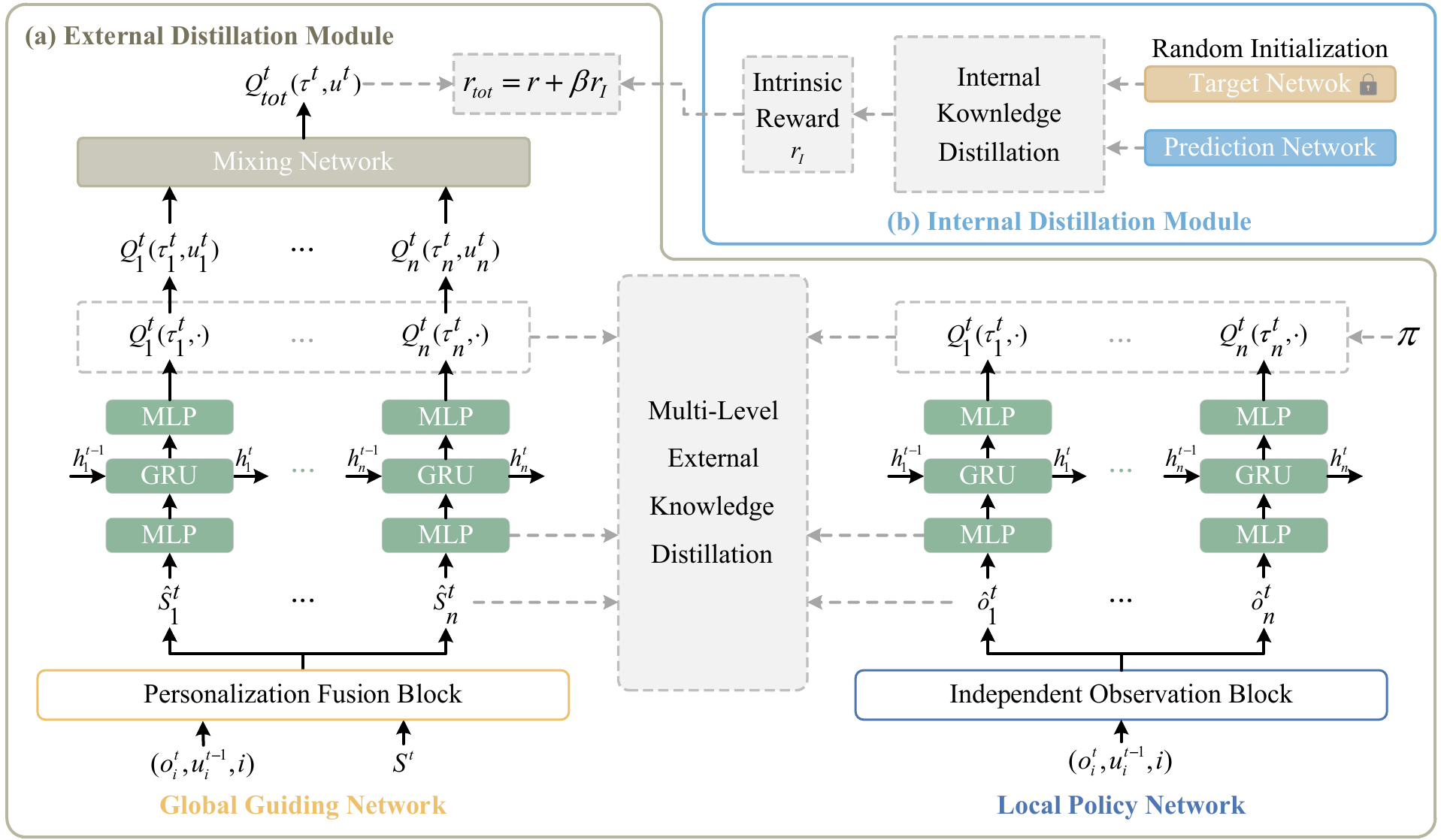}
    \caption{\textbf{The proposed DDN framework} consists of two parts: (a) the External Distillation Module, which includes the global guiding network (on the left) and the local policy network (on the lower right), and (b) the Internal Distillation Module (on the upper right).}
    \label{fig:DDN}
\end{figure*}

\begin{figure}[t]
    \centering
    \includegraphics[width=0.28\textwidth]{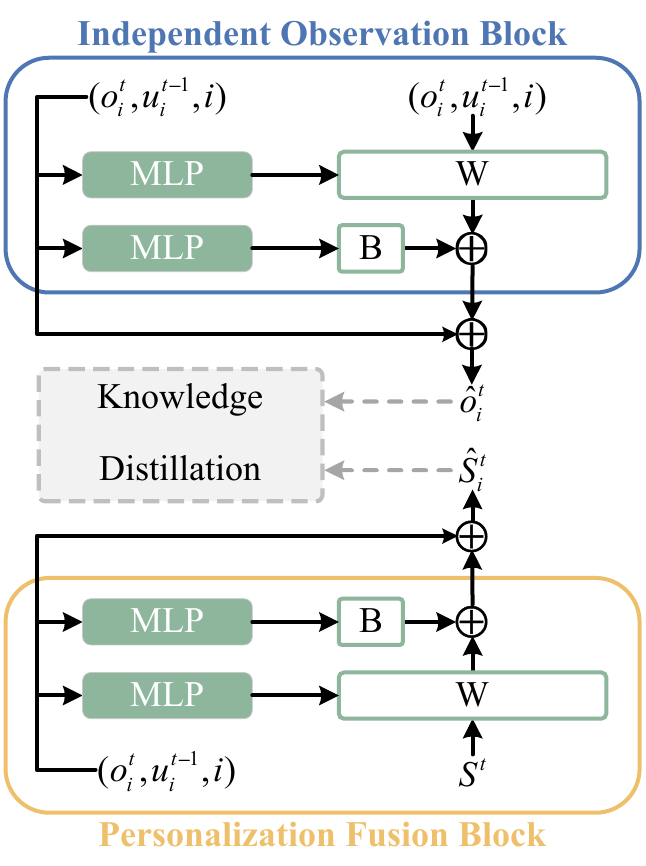}
    \caption{Knowledge distillation between the Personalization Fusion Block and the Independent Observation Block. }
    \label{fig:input}
\end{figure}

\section{Method}
In this section, we introduce the proposed DDN that comprises two main modules, shown in Figure \ref{fig:DDN}. The {\bf External Distillation Module} enhances agent performance by leveraging personalized global information and applying multi-level knowledge distillation for decentralized execution, reducing errors between centralized and local utility functions. The {\bf Internal Distillation Module} integrates global state information, generating intrinsic rewards to encourage exploration during training.

\subsection{Inherent Error and Knowledge Distillation}
During the training of agents, a limited field of view may not capture global changes. However, the influence of the global state is usually ignored in the process of value decomposition by existing CTDE methods. In other words, in the process of fitting the centralized value function, the utility functions with partial observation information are always lossy, which is re-expressed as:

\begin{equation}
\label{formula:IGM_inherent_error}
    \begin{aligned}
        Q_{tot}(s^{t}, \mathbf{u}^{t}) & = \mathcal{F}(Q_1^t(s^t_1,u^t_1),\dots,Q_n(s^t_n,u^t_n)) \\
        & \approx \mathcal{F}(Q_1^t(\tau^t_1,u^t_1),\dots,Q_n(\tau^t_n,u^t_n))\\
        & \approx r+\gamma \mathcal{F}(\max_{u^{t+1}_1}Q_1(\tau^{t+1}_1,u^{t+1}_1),\dots, \\
        & \max_{u^{t+1}_n}Q_n(\tau^{t+1}_n,u^{t+1}_n)).
    \end{aligned}
\end{equation}

This error will accumulate in the process of updating the $Q_{tot}$ function and eventually affect the training of the collaborative model~\cite{hong2022rethinking}. So far, the key to the problem is how to eliminate the inherent errors caused by inadequate fitting of global state information in the process of value decomposition. One potential approach is to complement the utility function with global state information, which is only available in the training phase as:

\begin{equation}
\label{formula:IGM_equ}
    \begin{aligned}
        Q_{tot}(s^{t}, \mathbf{u}^{t}) = \mathcal{F}(Q_1^t(s^{t},u^t_1),\dots,Q_n(s^{t},u^t_n))
    \end{aligned}
\end{equation}

Since both sides of the Equation \eqref{formula:IGM_equ} are fitted to the function by the global state, there is no lossy decomposition caused by insufficient observation. DDN designs a knowledge-distillation-based structure and then fits $Q_i^t(s^{t},u^t_i)$. The student network still uses $o_i$ as input, and then $Q_{stu\_i}^t(\tau_{stu\_i}^t,u_{stu\_i}^t)$ of student network is obtained from teacher $Q_{tea\_i}^t(s_{tea}^{t},u_{tea\_i})$ through knowledge distillation, thereby addressing the accumulated inherent error.

\subsection{External Distillation Module}
The External Distillation Module attempts to integrate personalized state information, and addresses the issue of information asymmetry during the training process of multi-agent systems through knowledge distillation, as shown in Part (a) of Figure \ref{fig:DDN}. The personalized state information is integrated into GGN to provide lossless knowledge. The LPN then performs distillation on the GGN based on local observations, transforming the execution process from being globally state-dependent to locally observation-dependent.

\subsubsection{Global Guiding Network}
Redundant global state information can impair agents' decision-making in multi-agent systems, which has been proven in previous work~\cite{yu2022surprising,ijcai2024p0004}. To address this, we propose a personalized approach to global state information, as shown in the Personalization Fusion Block in Figure \ref{fig:input}. This block uses the agent's local information $({o}_{i}^{t},u_{i}^{t-1},i)$ to generate weights $W$ and biases $B$, which vary across agents based on their observations. The global state information $S$ is then linearly transformed using $W$ and $B$ to produce personalized state information $\hat{S}_{i}^{t}$ for each agent, defined as $\hat{S}_{i}^{t}=S\times W+B$.

The personalized state $\hat{S}_{i}^{t}$ is processed by the agent network, which integrates a multi-layer perceptron (MLP) and a gated recurrent unit (GRU) to estimate each agent's individual action-value function $Q_i^t$. The mixing network then combines these individual $Q$-values into the joint action-value function $Q_{tot}^t$ using global state information. This design eliminates inherent errors between the centralized value function and local utility functions caused by limited observation fields.
Overall, the personalized state information enhanced GGN is able to estimate a better $Q$-function and guiding the agent policy network to update, with the global loss as:

\begin{equation}
    \label{formula:tdloss}
    L_{global}=\mathbb{E}[(r_{tot}+\gamma\cdot\max_{\boldsymbol{u}^{t+1}}Q^{-}_{tot}(\tau^{t+1},\boldsymbol{u}^{t+1})\\
    -Q_{tot}(s^{t}, \boldsymbol{u}^{t})]
\end{equation}

\noindent where $Q^-_{tot}$ denotes the target network. Its parameters are periodically synchronized with those of the current $Q$ network. This $r_{tot}$ refers to the total reward function with the intrinsic reward generated by the LPN, which is intended to directly leverage global state information to increase exploration, and its calculation is described in the following section.

\subsubsection{Local Policy Network}
To leverage global information during training while ensuring decentralized execution, the two blocks in Figure \ref{fig:input} share identical structures. Independent observational information $\hat{o}_{i}^{t}$ is defined as $\hat{o}_{i}^{t}=S\times W+B$. Notably, the LPN does not learn directly from the environment but leverages knowledge distillation, interacting with the GGN through multi-layer external distillation. The shared network structure ensures consistency in information flow between modules, streamlining the distillation process. Using only local observational information, the LPN gradually learns personalized global information, decision-making processes, and outcomes from the GGN, enabling decentralized execution. The multi-level external knowledge loss is formally computed as:

\begin{equation}
    \begin{aligned}
        L_{B} & =\frac{1}{n}\sum_{i=1}^{n}(\hat{o}_{i}^{t}-\hat{S}_{i}^{t})^{2} \\
        L_{Q} & =\sum_{i=1}^{n}Q_{i}^{GGN}(\tau_{i},\cdot)\log\frac{Q_{i}^{GGN}(\tau_{i},\cdot)}{Q_{i}^{LPN}(\tau_{i},\cdot)} \\
        L_{F} & =\sum_{i=1}^nF_i^{GGN}\log\frac{{F_i^{GGN}}}{F_i^{LPN}}
    \end{aligned}
\end{equation}%

\noindent where $L_B$ denotes the loss between the Personalization Fusion Block and the Independent Observation Block, $L_Q$ represents the $Q$-value loss for the LPN learning from the GGN, and $L_F$ refers to the intermediate feature loss from the MLP. The overall loss for the LPN is defined as:

\begin{equation}
\label{formula:KDloss}
    L_{local}=L_B+L_Q+L_F
\end{equation}%

\subsection{Internal Distillation Module}
\begin{figure}[t]
    \centering
    \includegraphics[width=0.35\textwidth]{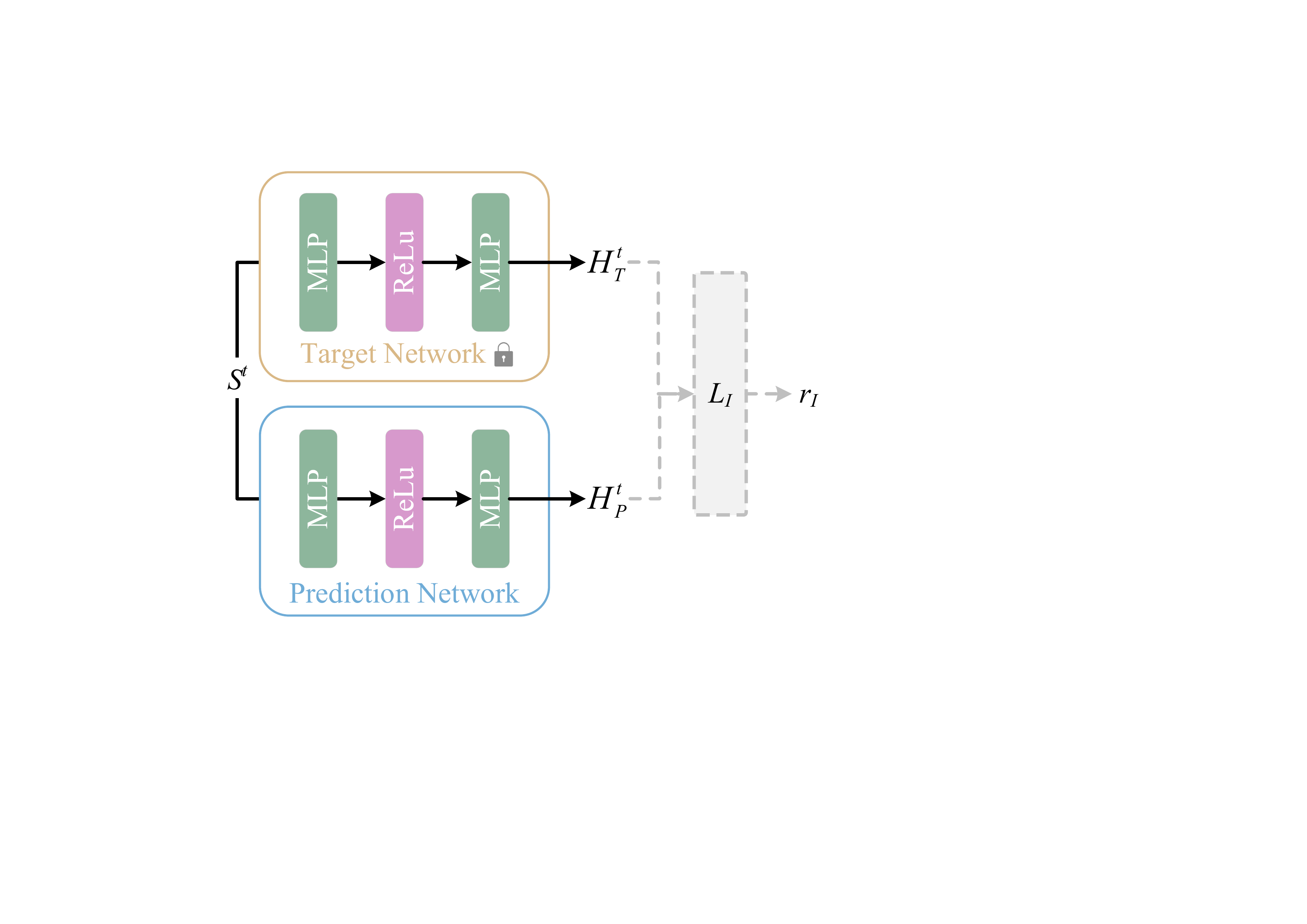}    
    \caption{Detailed outline of Internal Distillation Module.}
    \label{fig:IDM}
\end{figure}

To better utilize global state information, we propose an exploration method inspired by RND~\cite{burda2018exploration}, requiring fewer trials to identify useful actions and optimal policies. As illustrated in Figure \ref{fig:IDM}, the global state information $S^t$ is fed into both the target and prediction networks in the Internal Distillation Module (IDM) to produce $H_T^t$ and $H_P^t$, respectively. These networks share the same structure (MLP with ReLU), but with randomly initialized and fixed parameters in the target network. IDM employs the MSE loss to train the prediction network via distillation, with the prediction error updating the network and serving as intrinsic rewards $r_I$ to drive exploration. Higher prediction errors on novel or rarely seen states quantify their novelty, encouraging further exploration based on the agent's past experiences.

\begin{equation}
\label{formula:L_I}
    L_I=\frac{1}{n}\sum_{i=1}^{n}(H_P^t-H_T^t)
\end{equation}%

\begin{equation}
\label{formula:r_I}
    r_I=\mu L_I
\end{equation}%

\noindent Here, $\mu$ represents a random mask probability, ensuring that intrinsic rewards are based on a subset of experiences. This noise reduces reliance on existing information, introduces uncertainty, and diversifies the reward distribution in the IDM. Consequently, the total reward function of the DDN model in Equation \eqref{formula:tdloss} is expressed as:

\begin{equation}
    r_{tot} = r + r_I
\end{equation}

\subsection{The Overall Framework of DDN}
The DDN training process proceeds as follows: DDN computes intrinsic rewards efficiently via IDM to generate $r_{tot}$ and updates the GGN using $L_{global}$. Simultaneously, LPN performs multi-layer knowledge distillation to transfer GGN features with $L_{local}$, producing the utility function for agent execution. Meanwhile, the parameters of IDM are updated according to the Equation \eqref{formula:L_I}. Through the isolation design of the distillation structure, DDN can effectively eliminates the inherent errors caused by the decomposition of lossy values and improve the training efficiency of the collaborative policy. The Algorithm of DDN can be found in Section \ref{sup:A} of Supplementary Materials.

\begin{table}[b]
    \centering
    \begin{tabular}{c|c|c}
        \toprule
        Algorithm  & 3s\_vs\_5z & MMM2\\
        \midrule
        VDN          & 0.9089   & 0.0352 \\
        VDN\_DDN     & \textbf{0.9453}   & \textbf{0.3229}  \\
        \midrule
        Qatten       & 0.6510   & 0.6615 \\
        Qatten\_DDN  & \textbf{0.8945}   & \textbf{0.7422} \\
        \bottomrule
    \end{tabular}
    \caption{Performance of methods before and after integrating DDN.}
    \label{tab:vdn}
\end{table}

\section{Experiments}
In this section, we validate the effectiveness of the proposed method (DDN) by comparing it against several classic MARL baseline algorithms. These baseline algorithms include VDN, QMIX, QTRAN, Qatten, WQMIX, and QPLEX. All algorithms are implemented using the PyMARL framework~\cite{samvelyan2019starcraft} and evaluated on the StarCraft Multi-Agent Challenge (SMAC)~\cite{samvelyan2019starcraft} and the Predator-Prey environment. To ensure fairness, we align DDN’s hyperparameter settings as closely as possible with those of the baseline algorithms. Parameters such as learning rate, discount factor, and batch size are kept identical across all methods. Additionally, we use results from six independent runs to minimize the impact of randomness on the evaluations. The parameters of all compared experiments are shown in Section \ref{sup:B3} Supplementary Materials.

\begin{figure*}[t]
    \centering
    \includegraphics[width=0.9\textwidth]{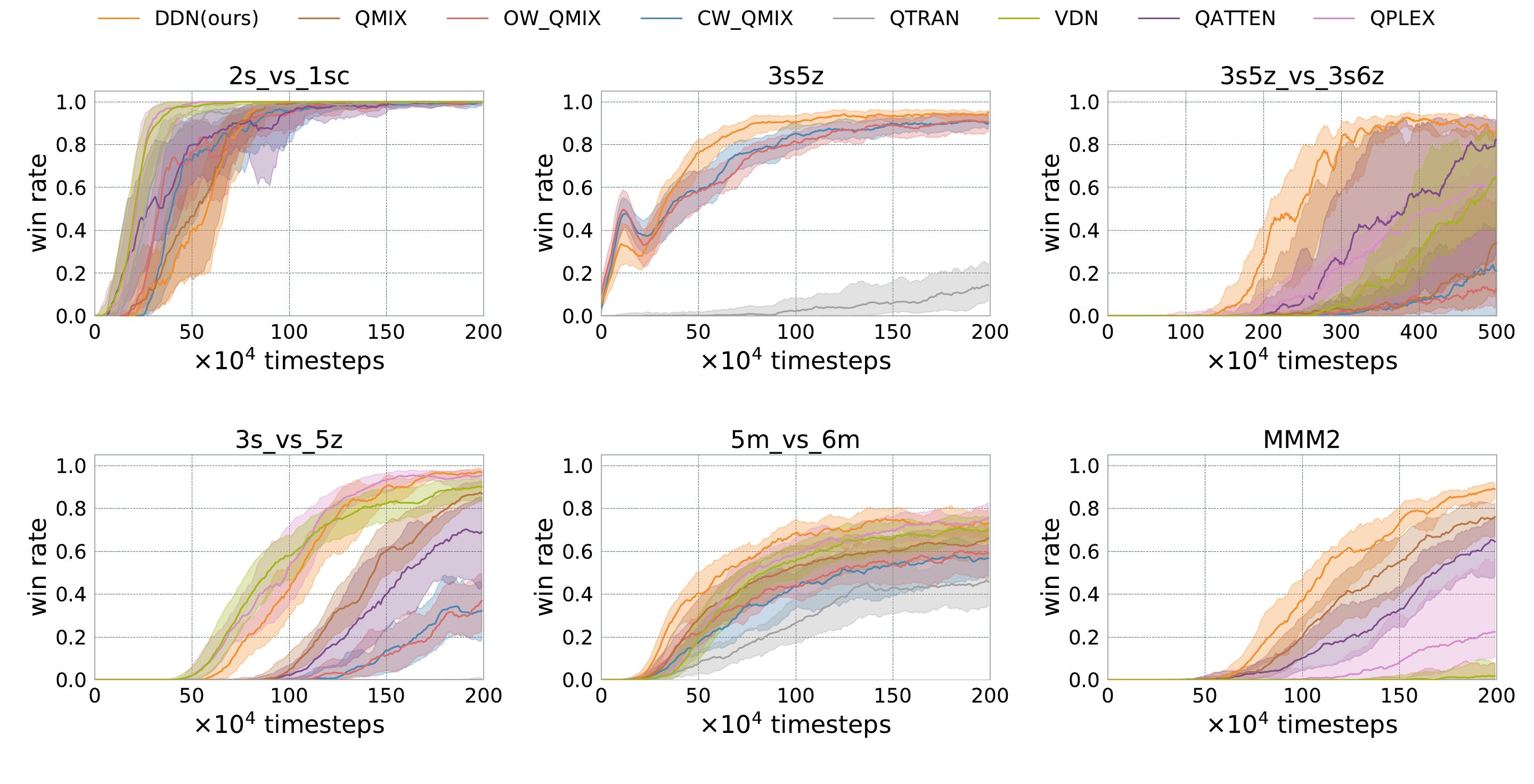}
    \caption{The win rates of different algorithms across the 6 combat scenarios in SMAC.}
    \label{fig:result}
\end{figure*}

\begin{table*}[t]
    \centering
    \begin{tabular}{c|c|c|c|c}
        \toprule
        Algorithms  & 3s\_vs\_5z & 5m\_vs\_6m & MMM2 & 3s5z\_vs\_3s6z \\
        \midrule
        IGMDA      & /              & 0.5048/0.3815    & 0.5607/0.1312  & / \\
        CTDS       & /              & 0.6690/0.4810    & 0.8060/0.2810  & / \\
        \midrule
        PTDE       & 0.9920/0.8870  & 0.8060/0.6900    & /              & 0.7760/0.6740 \\
        \midrule
        DDN(ours)  & 0.9701/0.9323  & 0.7188/0.5404    & 0.8789/0.6042  & 0.8776/0.4010 \\
        \bottomrule
    \end{tabular}
    \caption{Comparison of the results between the two phases. The win rate of (the global guiding network/the local policy network) for DDN, the win rate of (the teacher network/the student network) for PTDE and CTDS, and the win rate of (DAgger/behavioral cloning) for PTDE and CTDS. The symbol ``/" with no numerical value indicates that the algorithm is not implemented in the scenarios.}
    \label{tab:other}
\end{table*}

\subsection{SMAC}
We evaluate the performance of DDN on the StarCraft Multi-Agent Challenge (SMAC), a platform specifically designed for multi-agent cooperative research in StarCraft II. The primary evaluation metric is the win rate. Six combat scenarios are considered: easy scenarios (3s5z, 2s\_vs\_1sc), hard scenarios (3s\_vs\_5z, 5m\_vs\_6m), and super-hard scenarios (3s5z\_vs\_3s6z, MMM2), respectively. The AI difficulty level is set to ``super hard" (level 7) by default. Detailed information on each combat scenario is shown in Section \ref{sup:B1} Supplementary Materials.

First, we validate whether utilizing global state information within the proposed DDN framework benefits existing CTDE methods. By integrating the distillation structure of DDN into VDN and Qatten, we observe some significant performance improvements in Table \ref{tab:vdn}, demonstrating the effectiveness of DDN's direct processing and utilization of state information.

Second, We test the performance of the leader part of DDN with the existing value decomposition methods. As illustrated in Figure \ref{fig:result}, our proposed DDN outperforms or matches the baseline level in the vast majority of combat scenarios. The performance improvement is particularly significant in some super-hard scenarios, such as MMM2 and 3s5z\_vs\_3s6z, which place greater emphasis on inter-agent coordination. The advantage of DDN stems from its effective utilization of the distillation structure, which not only eliminates inherent error but also leverages state information more efficiently to generate intrinsic rewards, thereby enhancing overall performance.
However, in simple scenarios like 2s\_vs\_1sc, its performance slightly lags behind others, despite achieving a near-perfect win rate. This may be due to DDN's reliance on personalized state information, which increases computational complexity and prolongs training. Additionally, the Internal Distillation Module might cause redundant exploration in scenarios where optimal outcomes can be achieved without complex information.

Furthermore, we compare DDN with similarly structured algorithms, as shown in Table \ref{tab:other}, which lists win rates at the leader-follower stages. While PTDE benefits from parallel training advantages~\cite{hu2023rethinkingtheimplementation}, DDN still achieves comparable performance and outperforms IGM-DA and CTDS in both stages. Notably, the significant knowledge transfer rate (from 45.69\% to 96.10\%) demonstrates DDN's effective utilization of global state information and its ability to reduce inherent errors between the centralized value function and global state integration.

\subsection{Predator-Prey}
The Predator-Prey environment involves two roles: predator and prey. In this study, we consider 8 predators coordinating to capture 8 prey on a $10\times10$ grid, where each agent has a $5\times5$ sub-grid sight range. If two adjacent predators execute a capture action, the prey is caught, and the predators receive a reward $r = 10$. However, if a single predator attempts to capture a prey independently, it incurs a penalty of $p = 2$.

\begin{figure}[h]
    \centering
    \includegraphics[width=0.33\textwidth]{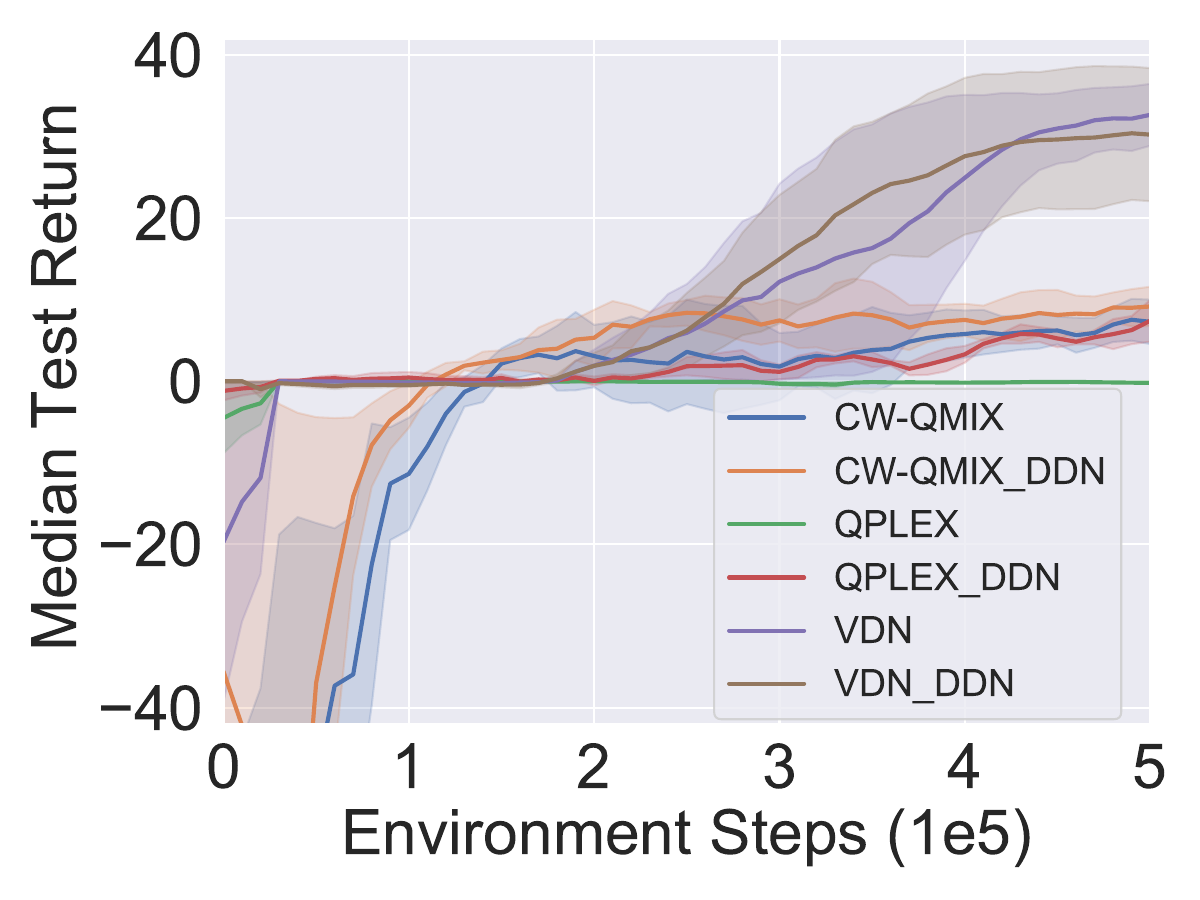}
    \caption{The comparing results on Predator-Prey.}
    \label{fig:pp_result}
\end{figure}

We evaluate the performance of DDN in the Predator-Prey environment by integrating it with VDN, QPLEX, and WQMIX. As shown in Figure \ref{fig:pp_result}, DDN demonstrates superior effectiveness compared to baseline algorithms in this environment, proving that the use of the two distillation modules facilitates better coordination among predators.

\begin{table}[b]
    \centering
    \begin{tabular}{c|c|c}
        \toprule
        Scenarios  & personalized state & raw state \\
        \midrule
        3s\_vs\_5z  & \textbf{0.9701} & 0.8216 \\
        \midrule
         MMM2       & \textbf{0.8789} & 0.1237 \\
        \bottomrule
    \end{tabular}
    \caption{Comparison between personalized $\&$ raw state in DDN.}
    \label{tab:state}
\end{table}

\subsection{Ablations}
In this subsection, we conduct ablation studies to investigate the impact of each component of the DDN. Two representative scenarios, 3s\_vs\_5z (hard) and MMM2 (very hard), are selected as the environments for this analysis.

To assess whether personalization improves DDN's use of global state information, we compare the decision-making outcomes of the Personalization Fusion Block with those using raw global state information. As shown in Table \ref{tab:state}, filtering global information through local observations to create personalized state information enables more efficient utilization of global data, leading to better strategies.

Then we test the effects of multi-level distillation versus single-level distillation on DDN by removing different loss components in Equation \eqref{formula:KDloss}, which is shown in Table \ref{tab:loss}. The results indicate that the win rate peaks when all three losses are applied together, highlighting the importance of the proposed Multi-Level External Knowledge Distillation.

\begin{table}[t]
    \centering
    \begin{tabular}{c|c|c|c|c}
        \toprule
        Scenarios  & $L_B$ & $L_Q$ & $L_F$ & Test Win Rates\\
        \midrule
        \multirow{2}{*}{3s\_vs\_5z}
            &                & $\checkmark$   &                & 0.9219  \\
            & $\checkmark$   & $\checkmark$   & $\checkmark$   & \textbf{0.9323}  \\
        \midrule
        \multirow{2}{*}{MMM2}
            &                & $\checkmark$   &                & 0.4258  \\
            & $\checkmark$   & $\checkmark$   & $\checkmark$   & \textbf{0.6042}  \\
        \bottomrule
    \end{tabular}
    \caption{The impact of different losses on the test win rate.}
    \label{tab:loss}
\end{table}

\begin{table}[t]
    \centering
    \begin{tabular}{c|c|c}
        \toprule
        Scenarios  & Probabilities & Test Win Rates\\
        \midrule
        \multirow{6}{*}{3s\_vs\_5z}  
            & \ding{55}     & 0.9518  \\
            & 0.10          & 0.9219  \\
            & 0.25          & 0.9036  \\
            & 0.50          & 0.9622  \\
            & \textbf{0.75} & \textbf{0.9701}  \\
            & 0.90          & 0.9219  \\
        \midrule
         \multirow{6}{*}{MMM2}
            & \ding{55}     & 0.7839  \\
            & 0.10          & 0.8542  \\
            & 0.25          & 0.7786  \\
            & 0.50          & 0.7773  \\
            & \textbf{0.75} & \textbf{0.8789}  \\
            & 0.90          & 0.8542  \\
        \bottomrule
    \end{tabular}
    \caption{The role of the Internal Distillation Module and the impact of mask probabilities on it. \ding{55} indicates that IDM is not used.}
    \label{tab:IDM}
\end{table}

% Additionally, Table 4 shows that using only $L_I$ achieves a higher win rate than using only $L_Q$, and even higher than using both $L_Q$ and $L_F$. This highlights the effectiveness of personalized state information and demonstrates that the Independent Observation Block's learning of Personalization Fusion Block knowledge helps to avoid the accumulation of inherent error.

Finally, we evaluate the effectiveness of the Internal Distillation Module by enabling or disabling it. Table \ref{tab:IDM} shows that win rates drop without IDM, confirming its positive impact on decision-making. Additionally, stochastic mask probabilities still affect IDM performance, with an optimal probability of $\mu=0.75$ enhancing state information utilization and promoting effective exploration.

\section{Conclusion}
In this paper, we propose a Double Distillation Network to eliminate the inherent error during the IGM condition satisfaction process. It effectively integrates global state information using a leader-follower framework and improves the training efficiency of the cooperative model. A multi-level distillation in the External Distillation Module is employed to eliminate the cumulative inherent error brought by the state-enabled centralized value function and the local utility functions. While Internal Distillation Module directly leverages state information to enhance the agents' curiosity for rarely encountered states, thereby promoting more effective exploration. Experiments on SMAC and Predator-Prey provide conclusive results to strongly validate the effectiveness and practicality of the proposed DDN framework.

\bibliographystyle{named}
\bibliography{ijcai25}

\clearpage
\appendix
\setcounter{table}{0}
\section{The Algorithm Implementation}
\label{sup:A}
The Double Distillation Network (DDN) consists of the External Distillation Module and the Internal Distillation Module. The former is responsible for reducing the continuously accumulated inherent error through distillation learning, while the latter makes full use of global information to generate environment-related intrinsic rewards. For better understanding, Algorithm \ref{alg:GGN} describes the detailed DDN implementation process.

\begin{algorithm}[hb]
    \caption{DDN algorithm}
    \label{alg:GGN}
    \textbf{Initialize}: replay memory $\mathcal{D}$, network with random parameters $\theta$, target parameters $\theta^-= \theta$, $step=0$.\\
    \mbox{\textbf{Initialize}: Observation $o=(o_1,\cdots,o_n)$ and state $S$.}
    \begin{algorithmic}[1] %[1] enables line numbers
        \WHILE{$step<step_{max}$}
            \STATE $t=0$, $s_0=\mathrm{initial \ state}$.
            \WHILE{$s^t\neq \mathrm{terminal}$ 
            nd $t<\mathrm{episode \ limit}$}
                \FOR{each agent $i$}
                    \STATE $\tau^{t}_{i}=\tau^{t-1}_{i}\cup\{(o^{t}_{i},u^{t-1}_{i})\}$;
                    \STATE With probability $\epsilon$ select a random action $u_t^i$;
                    \STATE Otherwise select $u_{i}^{t}=arg\max_{u_{i}^{t}}Q_{t}(\tau_{i}^{t},u_{i}^{t})$ for each agent $i$.
                \ENDFOR
                \STATE Execute action $u_i^t$ in environment and get reward $r^t$ and next state $S^{t+1}$;
                \STATE Set $t+1=t$ and $step=step+1$;
                \STATE Store transition $(\tau^t,\boldsymbol{u}^t,r^t,\tau^{t+1})$ in $\mathcal{D}$.
            \ENDWHILE
            \IF {$|D|>\mathrm{batch \ size}$}
                \STATE Sample random minibatch of transitions $(\tau^t,\boldsymbol{u}^t,r^t,\tau^{t+1})$ from $\mathcal{D}$ as $b$.
                \FOR{$t$ in each episode in batch $b$}
                    \STATE Set $step = step+1$;
                    \STATE Personalization fusion bolck's input with $S^t$ and $(o_i^t,u_i^{t-1},i)$ and obtain $\hat{S}_{i}^{t}$;
                    \STATE Independent observation block's input with $(o_i^t,u_i^{t-1},i)$ and obtain $\hat{o}_{i}^{t}$;
                    \STATE Calculate $Q$-value of global guiding network by $Q_{global}(S^{t},\mathbf{u}^{t})=IGM(Q_1^t(\hat{S}_{1}^{t},u^t_1),\dots,Q_n(\hat{S}_{n}^{t},u^t_n))$;
                    \STATE Internal distillation module's input with $S^t$ and obtain intrinsic reward $r_I$.
                \ENDFOR
                \STATE Update $Q_{global}$ by $L_{global}$;
                \STATE Update $Q$-value of local policy network by $L_{local}=L_B+L_Q+L_F$;
                \STATE Update internal distillation module by $L_I$.
            \ENDIF
            \IF {$step \% update-interval=0$}
                \STATE update the target network $\theta^-=\theta$.
            \ENDIF
        \ENDWHILE
    \end{algorithmic}
\end{algorithm}

\section{Experiment Overview}
\label{sup:B}
In this section, we introduce the DDN experimental platform, followed by a detailed description of the experimental setup and parameter configuration.

\subsection{SMAC Platform}
\label{sup:B1}
StarCraft II provides the StarCraft Multi-Agent Challenge (SMAC) platform for research on multi-agent collaboration. This platform is specifically designed to evaluate agents' abilities to learn cooperation and decision-making when solving complex tasks. Agents are required to learn coordinated combat strategies using reinforcement learning algorithms within a limited observation range. 
Each scenario simulates a confrontation between allied and enemy forces, where each allied unit is controlled by an independent agent. These agents must collaborate and strategize to defeat the enemy and secure victory. The scenarios are categorized into three levels of difficulty: easy, hard, and super-hard. In easy scenarios, allied forces generally have advantages in terms of numbers or unit types. In contrast, in hard and super-hard scenarios, allied forces often face significant numerical disadvantages or challenges arising from diverse enemy unit types. This setup effectively tests the collaborative capabilities of multi-agent systems in complex and asymmetric environments.
Table \ref{tab:SMAC} provides detailed information on each combat scenario used by DDN.
\begin{table*}[ht]
    \centering
    \begin{tabular}{lll}
        \toprule
        Scenario  & Ally Units & Enemy Units\\
        \midrule
        3s5z             & 3 Stalkers \& 5 Zealots  & 3 Stalkers \& 5 Zealots \\
        2s\_vs\_1sc      & 2 Stalkers               & 1 Spine Crawler  \\
        3s\_vs\_5z       & 3 Stalkers               & 5 Zealots \\
        5m\_vs\_6m       & 5 Marines                & 6 Marines \\
        3s5z\_vs\_3s6z   & 3 Stalkers \& 5 Zealots  & 3 Stalkers \& 6 Zealots \\
        MMM2             & 1 Medivac, 2 Marauders \& 7 Marines  & 1 Medivac, 2 Marauders \& 8 Marines \\
        \bottomrule
    \end{tabular}
    \caption{Description of the agent composition in combat scenarios.}
    \label{tab:SMAC}
\end{table*}

\subsection{Predator-Prey}
\label{sup:B2}
In multi-agent reinforcement learning, the Predator-Prey environment simulates the interactions between predators and prey, and is used to study cooperation, competition, and strategy selection among multiple agents. The Predator-Prey environment in DDN is implemented based on OpenAI's reinforcement learning framework, using a $10\times10$ grid world created by Gymnasium, where the predator's observation range is $5\times 5$. We consider 8 predators coordinating to capture 8 prey. Predators receive a reward of $r=10$ for collaborating to catch prey. However, if a single predator attempts to capture prey alone, it is penalized with a penalty value of $p=2$.

\subsection{Experimental Setup}
\label{sup:B3}
To ensure fairness, all algorithms are implemented based on the PyMARL framework~\cite{samvelyan2019starcraft}, and the hyperparameters of DDN are set to be consistent with the baseline algorithms as much as possible. The learning rate for the neural networks is uniformly set to $5\times {10}^{-4}$. The Global Guiding Network uses the RMSprop optimizer with a decay rate of 0.99, while the Local Policy Network employs the Adam optimizer. The discount factor $\gamma$ is fixed at 0.99 for all tasks and scenarios. Each agent independently applies an $\epsilon$-greedy policy for action selection. $\epsilon$ anneals from 1.0 to 0.05 over 50,000 time steps and remains fixed for the remaining training. Training runs for 2,000,000 time steps on the SMAC platform. For the 3s5z\_vs\_3s6z scenario, the learning rate is set to 0.001, and training runs for 5,000,000 time steps on the SMAC platform. The $\epsilon$ annealing time is extended from 50,000 to 1,000,000 time steps. In the Predator-Prey environment, training runs for 1,000,000 time steps, with each episode limited to 200 time steps.

\end{document}